\documentclass[12pt, a4paper]{article}
\usepackage[font=footnotesize]{caption}
\usepackage[utf8]{inputenc}
\usepackage{graphicx}
\usepackage{authblk}
\usepackage{booktabs}
\usepackage{verbatim}
\usepackage{url}
\usepackage{caption}
\usepackage{subcaption}
\usepackage{indentfirst}
\usepackage{url}
\usepackage{amssymb}
\usepackage{amsmath}
\usepackage{tcolorbox}
\title{\emph{Simple} Games on \emph{Complex} Networks}

\author{Alexandre Benatti$^1$ \\ Luciano da F. Costa$^2$}

\affil{
$^1$Institute of Mathematics and Statistics - DCC \\
University of S\~ao Paulo \\
Rua do Mat\~ao, 1010, \\ S\~ao Paulo, SP 05508-090 Brazil 
\\ \vspace{0.5cm}
$^2$S\~ao Carlos Institute of Physics - DFCM \\
University of S\~ao Paulo \\
Av. Trabalhador S\~ao-carlense, 400, \\ S\~ao Carlos, SP 13566-590 Brazil
}

\date{\emph{22th April, 2024}}

\begin{document}

\maketitle

\begin{abstract}
The relationship between topology and dynamics of complex systems has motivated continuing interest from the scientific community. In the present work, we address this interesting topic from the perspective of simple games, involving two teams playing according to a small set of simple rules, taking place on four types of complex networks. Starting from a minimalist game, characterized by full symmetry always leading to ties, four other games are described in progressive order of complexity, taking into account the presence of neighbors as well as strategies. Each of these five games, as well as their specific changes when implemented in four types of networks, are studied in terms of statistics of the total duration of the game as well as the number of victories and ties, with several interesting results that substantiate, in some cases, the importance of the network topology on the respective dynamics. As a subsidiary result, the visualization of relationships between the data elements in terms of coincidence similarity networks allowed a more complete and direct interpretation of the obtained results.
\end{abstract}

\section{Introduction}\label{sec:introduction}

The specific interconnecting (topologic) properties of a complex system can influence, or even determine, diverse types of dynamics taking place on that system. This important relationship between topology and dynamics has motivated continuing interest also in the area of Network Science (e.g.~\cite{barabasi2013network,newman2018networks,costa2007characterization,costa2011analyzing}).

Among the several possible types of dynamics taking place on complex networks, games involving interactions between two or more teams has also motivated interest from the scientific community developing studies on the relationship between topology and dynamics. Several works have focused on the prisoner dilemma (e.g.~\cite{perc2006double,poncela2007robustness,szolnoki2008diversity,cardillo2010co,konno2011condition}). Other developments considering diverse types of games taking place on complex networks include but are by no means limited to~\cite{dall2006nonequilibrium,sinatra2009ultimatum,gomez2011disentangling,li2015evolutionary,huang2018effects}.

The present work aims at studying in a systematic and incremental manner a sequence of basic games between two teams (A and B), in the sense of involving a small and simple set of rules, which can take place on diverse types of complex networks. More specifically, two take randomly selected individuals (one from each team) perform simultaneous choices at each successive time step or iteration. Simultaneity is of particular interest because it allows for complete symmetry to be achieved, always leading to ties between the two teams (all individuals from both teams vanishing). Also of interest in the present work is the topic of \emph{symmetry breaking}~\cite{strocchi2005symmetry,garlaschelli2010complex,mainzer2013local}, which is intrinsically related to dynamics in complex networks.

In order to investigate the possible impact of the network topology on the obtained game dynamics, we consider four types of complex networks, namely: (a) regular lattice (REG); (b) ER; (c) BA; and (d) GEO. These types of networks account for a good level of topological diversity, with the models REG and GEO not following the small-world property, while the BA model presents intense degree asymmetry (scale-free distribution of node degrees). The ER type can be understood as a stochastic version of the model REG.

Though the reported developments follow a mostly abstract perspective, which has been here identified as `\emph{games}', the considered types of dynamic and topology are related to several important real-world types of systems and problems, including but by no means being limited to interaction between particles~\cite{deen2007review,liu2019smoothed}, chemical reactions~\cite{maini1997spatial,andrews2004stochastic}, cell interactions~\cite{matsuo2012bone,taylor2012pairwise}, living beings~\cite{kareiva1988spatial,andersen1992spatial}, economical groups~\cite{zhang1998modeling,pushkin2004bank,li2016models}, among several other possibilities.

We start with what is possibly the simplest set of rules (game G1), involving uniformly choice among the available positions unoccupied by the adversary. Whenever two individuals meet, they cancel one another. Because of the complete underlying \emph{symmetry} of the choices, this type of game necessarily leads to ties, with the complete vanishing of both teams. Yet, despite this intrinsic simplicity, the total time taken by the game may still vary as a consequence of the specific topological features of each type of network, an aspect that is respectively studied in the present work respectively to each of the considered types of game and network topology.

In order to break the symmetry of case G1, the rules are changed so that whenever two individuals meet, the one having the largest number of neighbors from the same team prevails, so that the adversary is removed from the game. This second type of game is referred to as G2. Now, victories by any of the two teams become possible, as well as ties, so that it becomes interesting to consider the respective statistics.

The third type of game (G3) studied in the present work differs importantly from the two previous cases in the aspect that it incorporates a \emph{strategy} by one of the teams, while the other makes random choices as in G1 and G2. More specifically, the current player of team A (randomly selected among the players of that team) moves \emph{deterministically} to the node containing the largest number of neighbors from its own team. In this particular case, two more nodes have the same number of neighbors, and one of them is taken with uniform probability. It is expected that team A will have more chance to win as a consequence of adopting the above described strategy.

Another type of game (G4) is also considered here, characterized by both teams (and not only team A) adopting the above described \emph{deterministic} strategy. Now, it could be expected that the individuals of each team would tend to remain adjacent one another, irrespectively of the individuals of the other team.

The last type of game (G5) studied in the present work consists in a modification of G4 in which each choice is taken in \emph{randomly} preferential manner, instead of deterministically as in the previous case. More specifically, each individual move randomly to the adjacent node preferentially to the number of neighbors of its own team. One particularly interesting issue related to G5 consists of the possible effects of the random preferential choice leading to greater interaction between the two teams.

Table~\ref{tab:games} provides, for the sake of quick reference, an abbreviated description of the five types of games considered in the present work.

\begin{table}
\centering
\caption{Abbreviated description of the five types of games considered in the present work.}\label{tab:games}
\begin{tabular}{|c|c|}
\hline
\textbf{Game} & \textbf{Description} \\ \hline
\textbf{G1}   &  Uniformly random choice of next positions.    \\ \hline
\textbf{G2}   &  As G1, but the player with more neighbors prevails.   \\ \hline
\textbf{G3}   &  \begin{tabular}[c]{@{}c@{}}As G2, but one of the teams moves to the \\ position with the largest number of neighbors. \end{tabular}                   \\ \hline
\textbf{G4}   & \begin{tabular}[c]{@{}c@{}} As G2, but both teams move deterministically to the \\ position with the largest number of neighbors.\end{tabular}                   \\ \hline
\textbf{G5}   & As G4, but both teams move in randomly preferential manner.                    \\ \hline
\end{tabular}
\end{table}

A subsidiary development described in the present work concerns the use of coincidence similarity networks as a means to obtain more complete and intuitive visualization and analysis of the relationships between each of the possible considered combinations of types of games and types of network topology.

This work starts by reviewing the main adopted concepts and methods, then presenting and discussing the performed experiments and obtained results respectively to percentages of victories/ties as well as the duration of the games. The conclusions then provide a summary of the main findings as well as some prospects for future developments.

\section{Methodology}\label{sec:concepts}

In this study, four distinct network topologies were employed to run the execution of the presented games. 

The first network was a simple regular lattice (REG), in 2 dimensions, with each node connecting to its nearest neighbors. Another topology employed was an Erdős–Rényi~\cite{erdos1959random} (ER). We also use Barabási–Albert~\cite{barabasi1999BA} (BA) networks. The last type of topology adopted was a geometric network (GEO), corresponding to the Delaunay triangulation (e.g.~\cite{riedinger1988delaunay}) of a slightly perturbed version of an orthogonal lattice.

Given a dataset of $N$ entities, each with $M$ respective properties, characteristics, or \emph{features}, it is often interesting to \emph{standardize} (e.g.~\cite{gewers2018principal}) the values of the latter, which can be implemented by using the following expression:
\begin{align}
   \tilde{X_i^j} = \frac{X_i^j - \mu_{j}}{\sigma_{j}}
\end{align}

with $i = 1, 2, \ldots, N$ and $j = 1, 2, \ldots, M$. In addition, $X_i$ stands for the feature $j$ of each respective sample $i$, and $ \mu_{j}$ and $ \sigma_{j}$ correspond, respectively, to the average and standard deviation of feature $j$ among all the $N$ entities.

Standardizing a specific dataset assumes that the range of magnitudes of the features involved in that dataset are not important, as they will all be normalized to unit standard deviation as a consequence of the standardization procedure. In addition, the average value of each of the features should also be known to have no importance for subsequent analysis of the dataset.
Indeed, the standardization of a dataset implies that the normalized data to have null means and unit standard deviation. In addition, the feature values of most of the entities will be comprised between $-2$ and $2$. The standardized values are non-dimensional. One of the interesting characteristics of data standardization is that all features will have similar potential influence on the respective analysis, even in cases involving features with markedly distinct extreme values. 

Though several approaches can be employed for data analysis, the consideration of similarity indices provides some interesting intrinsic characteristics. In particular, we have the possibility to express the relationship between pairs of data elements in terms of the similarity between their respective features, therefore providing an intrinsically intuitive interpretation. As described recently, the coincidence similarity index (e.g.~\cite{costa2021similarity,da2021further,da2021multiset,da2022coincidence}) can be applied as a means to perform strict comparison between real-valued data elements, being also normalized and presenting good tolerance to outliers and data noise. 

With the help of multiset concepts (e.g.~\cite{da2021multisets}), and particularly by using the extension to negative values described in~\cite{costa2023mulsetions}, the Jaccard index (e.g.~\cite{Jaccard1, Loet, Jac:wiki, da2021further}) involved in the coincidence similarity calculation can be applied to two non-zero real-valued vectors $\vec{v}$ and $\vec{r}$ as:
\begin{align}\label{eq:jaccard}
    \mathcal{J}(\vec{v}, \vec{r}) = \frac{\sum_i \{ \text{min}(v_i^P,r_i^P) + \text{min}(|v_i^N|,|r_i^N|) \} }{\sum_i \{ \text{max}(v_i^P,r_i^P) + \text{max}(|v_i^N|,|r_i^N|)  \}}
\end{align}

where $\vec{v}^P$ and $\vec{v}^N$ are the vectors with the positive and negative values of $\vec{v}$.

We have that $0 \leq \mathcal{J}(\vec{v}, \vec{r})  \leq 1$ and $\mathcal{J}(\vec{v}, \vec{r}) = \mathcal{J}(\vec{r}, \vec{v})$. The two vectors are maximally similar at $\mathcal{J}(\vec{v}, \vec{r}) = 1$ and maximally dissimilar for $\mathcal{J}(\vec{v}, \vec{r}) = -1$.

As an example of numeric calculation, consider the two following vectors $\vec{v}$ = [1.5,-2.1] and $\vec{r}$ = [3.5,-1.6]. It follows that $\vec{v}^P$ = [1.5,0], $\vec{r}^P$ = [3.5,0], $\vec{v}^N$ = [0,-2.1], and $\vec{r}^N$ = [0,-1.6].

We have from Equation~\ref{eq:jaccard} that the Jaccard index can then be calculated as: 
\begin{align}
    \mathcal{J}(\vec{v}, \vec{r}) = & \frac{[\text{min}(1.5,3.5) + \text{min}(0,0)] + [\text{min}(0,0) + \text{min}(2.1,1.6)]}{[\text{max}(1.5,3.5) + \text{max}(0,0)] + [\text{max}(0,0) + \text{max}(2.1,1.6)] } = \nonumber \\
    & =  \frac{1.5 + 1.6}{3.5 + 2.1} = \frac{3.1}{5.6} \approx 0.5536  \nonumber
\end{align}

The extension of the interiority index (also called overlap, e.g.~\cite{vijaymeena2016a}) to include negative values can be expressed as:
\begin{align}\label{eq:interiority}
    \mathcal{I}(\vec{v}, \vec{r}) = \frac{\sum_i \{ \text{min}(v_i^P,r_i^P) + \text{min}(|v_i^N|,|r_i^N|) \} }{\text{min}( \sum_i |v_i| , \sum_i |r_i| )}
\end{align}

We also have that $0 \leq \mathcal{I}(\vec{v}, \vec{r})  \leq 1$ and $\mathcal{I}(\vec{v}, \vec{r}) = \mathcal{I}(\vec{r}, \vec{v})$.

Because the Jaccard similarity index does not take into account the relative interiority between the two compared multisets (or vectors), the \emph{coincidence similarity index} between to non-zero vectors with possibly negative entries has been defined as follows:
\begin{align}\label{eq:coincidence}
    \mathcal{C}(\vec{v}, \vec{r}) =  \mathcal{J}(\vec{v}, \vec{r})^D \,  \mathcal{I}(\vec{v}, \vec{r})
\end{align}

where $D$ is a parameter controlling how strict the performed similarity comparison is performed.  More strict comparisons are obtained for larger values of $D$.

Expression~\ref{eq:coincidence} tends to provide maximally strict comparisons when applied over a dataset that has been previously standardized. That is because the data elements then become distributed around the center of the features coordinate system, where the coincidence similarity is most sensitive (e.g.~\cite{da2021multiset}). At the same time, as the result of standardization one or more possible clusters may fall near the center of coordinate, becoming less likely or even impossible to be detected. In these cases, it is possible to consider returning the dataset center of mass to its original position, which can be done by standardization followed by adding the respective features average. Data normalized in this manner will have similar magnitude variations around the original center of mass. In the particular case in which the original dataset already had center of mass near the origin of the coordinate system, a same positive constant can be added to each of the involved features, therefore displacing the dataset away from the coordinates center. Another possibility to control the sensitivity of the similarity comparisons near the center of the features coordinates consists of regularizing expression~\ref{eq:jaccard} by adding a same positive constant value to both its numerator and denominator.

As described in~\cite{da2022coincidence}, \emph{coincidence similarity networks} can be readily obtained from the respective coincidence similarities among the elements of the dataset. More specifically, each node corresponds to one of the data elements (or entries), while the links between pairs of nodes have weights equal to the respective pairwise coincidence similarity.

\section{Results and Discussion}\label{sec:results}

The five types of games considered in the present work combined with the four types of complex networks lead to 20 cases to be studied in the present section. First, we investigate how these cases compare one another regarding the respective statistics of victories and ties. Subsequently, the 20 cases are compared while taking into account the respective histogram of total duration in basic steps.

All networks have 25 nodes and similar average degrees (REG: 5.76, ER: 5, BA: 3.76, and GEO: 5.08). All games are performed 10,000 times. A total of 10 players for each team are initially distributed among the network nodes in uniformly random manner. In all games, the active player can move to a node that is either empty or has one or more players from the same team. All network visualizations depicted in the present work adoptd the Fruchterman-Reingold methodology (e.g.~\cite{fruchterman1991graph}).

The percentages of victories and ties obtained for each of the 20 cases are depicted in Table~\ref{tab:results}, presented as $\left[p_A, t, p_b \right]$, where $p_A$ and $p_B$ correspond to the percentages of victories by teams A and B, and $p_t$ indicates the percentage of ties.

\begin{table}
\small
\centering
\caption{The statistics (percentages) of victories and ties respective to each of the 20 considered
combinations of game and network topology types. The asterisk indicates that the games never end.}\label{tab:results}
\begin{tabular}{|c|c|c|c|c|}
\hline
\multicolumn{1}{|c|}{\textbf{}}      & \textbf{REG} & \multicolumn{1}{c|}{\textbf{ER}} & \multicolumn{1}{c|}{\textbf{BA}} & \multicolumn{1}{c|}{\textbf{VR}} \\ \hline
\multicolumn{1}{|c|}{\textbf{G1}} & {[}0, 100, 0{]}         & {[}0, 100, 0{]}                           & {[}0, 100, 0{]}                           & {[}0, 100, 0{]}                           \\ \hline
\multicolumn{1}{|c|}{\textbf{G2}} & {[}44.4, 10.4, 45.2{]}         & {[}43.0, 14.5, 42.5{]}                           & {[}46.3,  9.8, 43.9{]}                           & {[}42.1, 14.3, 43.6{]}                           \\ \hline
\textbf{G3}                       & {[}100, 0, 0{]}         & {[}99.3,  0.1,  0.6{]}                           & {[}96.1,  0.9, 3.0{]}                           & {[}100, 0, 0{]}                           \\ \hline
\textbf{G4}                       & {[}0,100, 0{]}$^*$         & {[}44.8, 10.3, 44.9{]}                           & {[}44.8, 10.6, 44.6{]}                           & {[}44.4, 10.9, 44.7{]}                           \\ \hline
\textbf{G5}                       & {[}46.5,  6.4, 47.1{]}         & {[}45.6,  9.3, 45.1{]}                           & {[}46.8,  6.8, 46.4]{]}                           & {[}45.8,  8.9, 45.3{]}                           \\ \hline
\end{tabular}
\end{table}

A preliminary analysis of Table~\ref{tab:results} indicates a relatively minor influence of the topologies REG, ER, BA, and GEO on the results. A relatively small probability of tie characterizes the results obtained for these cases, while the percentages of victory are substantially higher. However, markedly distinct results have been obtained for the REG topology respectively to G3 and G4.

\begin{figure}
  \centering
     \includegraphics[width=1 \textwidth]{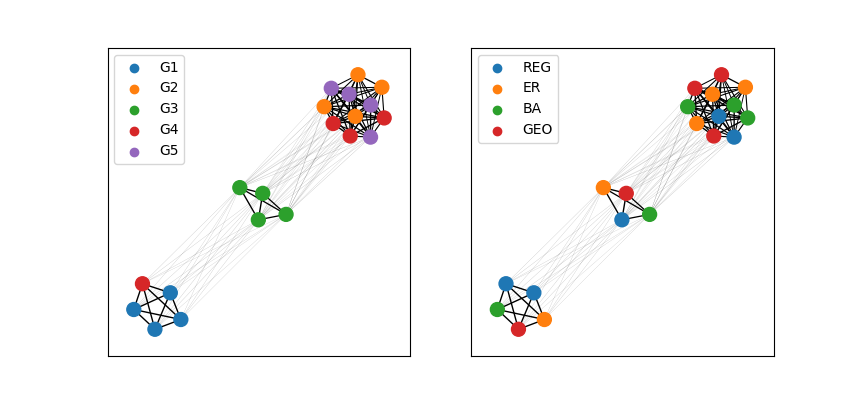} \\
     \hspace{.5cm} (a) \hspace{4.7cm} (b) \\
   \caption{The coincidence similarity networks, derived from Table~\ref{tab:results} and using $D=1$, expressing the relationships among the 20 considered configurations of game/type of network labeled according to game types (a) and complex network models (b).}\label{fig:sim_net1}
\end{figure}

The analysis of tabulated data can be greatly enhanced by considering the respective coincidence similarity network, which is shown in Figure~\ref{fig:sim_net1} relative to labeling the nodes (cases) by type of game (a) and type of network topology (b).

The network in (a) indicates in an effective manner that the types of games cluster neatly into three groups respectively to G1 and one element of G4, G3, as well as a larger group corresponding to the remainder of the configurations. The items within each of these three groups are intensely similar one another while being markedly distinct from the configurations in the other groups. We also have from the coincidence network shown in (b) that the network types do not influence the obtained clusters.

\begin{figure}
  \centering
     \includegraphics[width=.4 \textwidth]{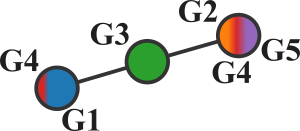}   \vspace{0.7cm} \\
   \caption{Summarization of the relative interconnections among the three clusters identified in Fig~\ref{fig:sim_net1}(a) as a respective coincidence complex network. The intermediate group G3 consists precisely to the game G3 implemented in the four considered topologies.}\label{fig:interconnection1}
\end{figure}

Figure~\ref{fig:interconnection1} presents the network in Figure~\ref{fig:sim_net1}(a) summarized in three groups corresponding to the detected clusters. The relationship between these clusters, which are associated to the types of games, indicates in a direct manner that the group G3 intermediates, in the sense of the percentages of victories/ties, the two other obtained groups.
 
Thus, as far as the statistics of victories and ties are concerned, the 20 considered cases resulted mostly clustered into three main groups respective to G1 and G3, with the remainder cases constituting the third cluster.

\begin{figure}[!ht]
  \centering
    \includegraphics[width=.99\textwidth]{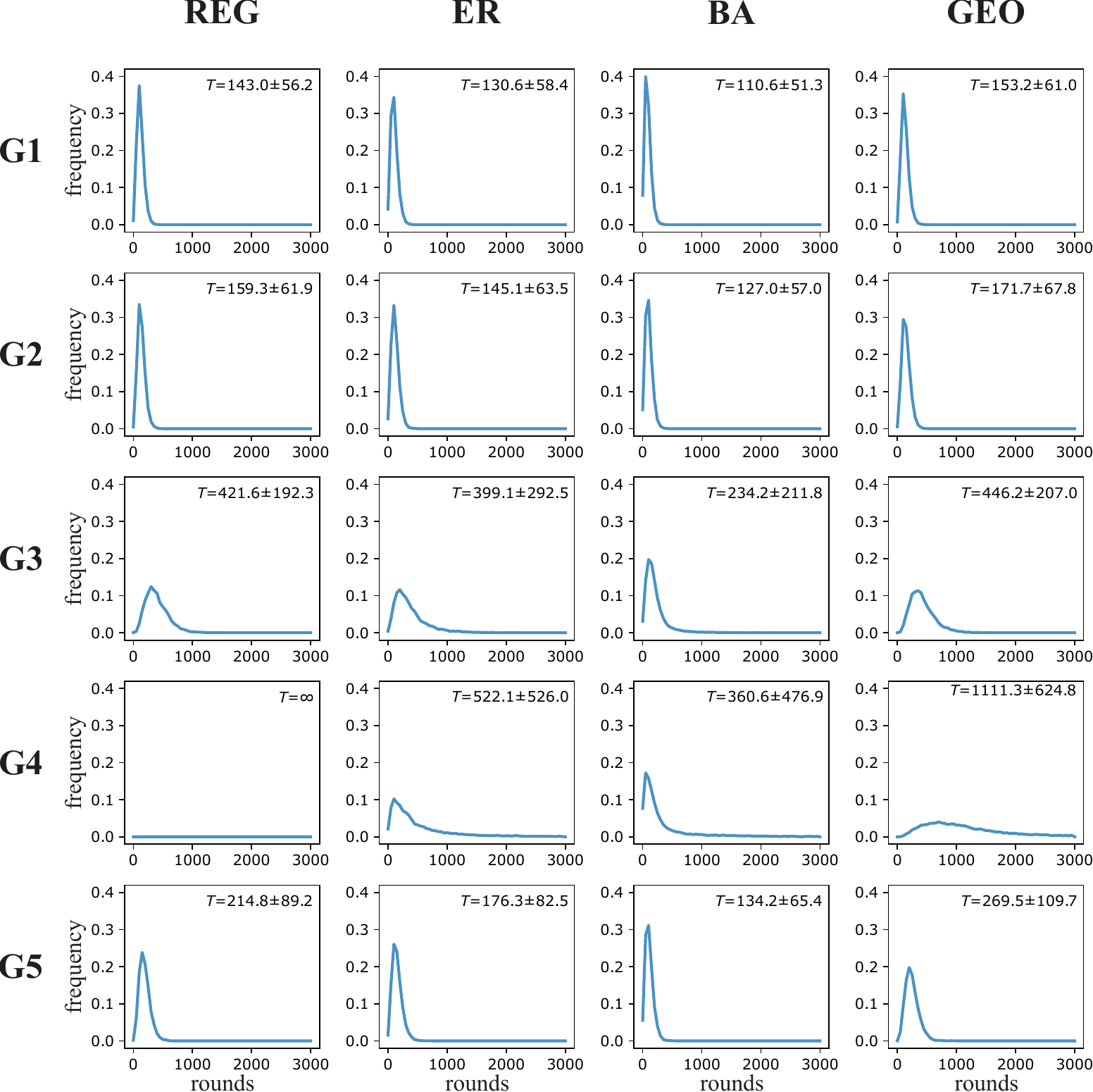}
  \caption {The histograms of the duration of the games respectively to the 20 considered cases. The insets indicate the average $\pm$ the standard deviation of the duration times.}
  \label{fig:times}
\end{figure}

Figure~\ref{fig:times} presents the histograms of duration times respectively to the 20 possible combinations of types of games and types of networks considered in the present work. The percentages of vitories/ties and duration times have been standardize prior to the calculation of the respective coincidence similarity networks. In addition, it should be observed that the case corresponding to G4 on the REG topology has not been considered because of the infinite duration of games obtained for this configuration.

\begin{figure}
  \centering
     \includegraphics[width=1 \textwidth]{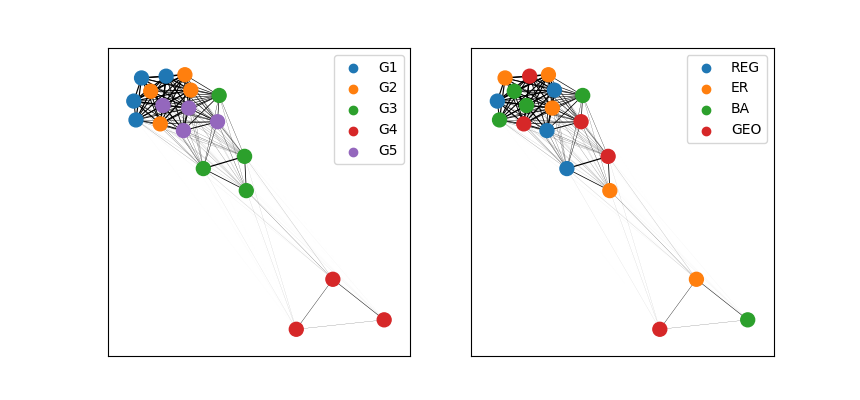} \\
     \hspace{.5cm} (a) \hspace{4.7cm} (b) \\
   \caption{The coincidence similarity network ($D=1$) obtained respectively to the duration of the games taking place in the several considered network topologies with the nodes shown in colors corresponding to the types of games (a) and types of networks (b).  In the former case, the types of game G1 led to a well-separated group of nodes, with the nodes corresponding to the type G3 intermediating the connections with the other three types of games.}\label{fig:sim_net2}
\end{figure}

The results in Figure~\ref{fig:times} indicate that the types of networks do not tend to influence substantially the game duration, except for the cases G3 and G4, in which distinct histograms have been obtained for the BA and GEO topologies. A more complete and systematic analysis of the histograms depicted in Figure~\ref{fig:times} can be achieve by considering the respective coincidence similarity network, which is shown in Figure~\ref{fig:sim_net2}.

\begin{figure}
  \centering
     \includegraphics[width=1 \textwidth]{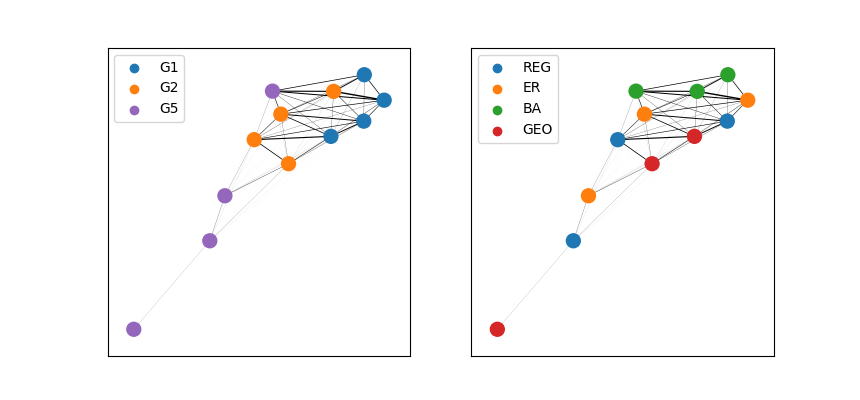} \\
     \hspace{.5cm} (a) \hspace{4.7cm} (b) \\
   \caption{The coincidence similarity network ($D=2$) considering the duration of only the games G1, G2, and G5 shown in colors corresponding to the types of games (a) and types of networks (b). Two cases of the game G5 resulted more substantially separated from the remainder of the network.}\label{fig:sim_net3}
\end{figure}

In order to minimize the effect of the marked difference among the games G3 and G4 and the other games on the visualization of the respective coincidence similarity network shown in Figure~\ref{fig:sim_net3}, a new coincidence similarity network disconsidering these two groups has been obtained, presented in Figure~\ref{fig:sim_net2}.

\begin{figure}
  \centering
     \includegraphics[width=.5 \textwidth]{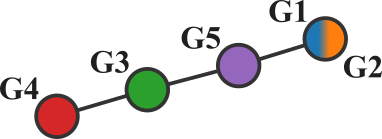}   \vspace{0.7cm} \\
   \caption{Summarization of the relative interconnections among the durations of the five types of games revealed by the coincidence similarity networks shown in Figs.~\ref{fig:sim_net2} and~\ref{fig:sim_net3}. The games G1 and G2 have short and mostly indistinguishable durations, with G4 leading the longest durations. The types of games G3 and G5 resulted as intermediating the two extreme groups G4 and G1/G2. }\label{fig:interconnection2}
\end{figure}

We have from Figures~\ref{fig:sim_net2} and~\ref{fig:sim_net3} that the types of games G1 and G2 are most similar among possible combinations when considered from the perspective of game duration, followed by G5. The types of games G3 and G4 resulted not only dissimilar to the other three types of games but also mostly different one another. We can also infer from the visualizations in Figures~\ref{fig:sim_net2} and~\ref{fig:sim_net3} that the game G4 is the most external (outliers) to the overall network, with the types of games G3 and G5 successively intermediating the connections with the remainder two types G1 and G2, which are mostly indistinct one another. This just discussed overall relationship between the five types of considered games, which has been allowed by the consideration of the coincidence similarity network, is summarized in Figure~\ref{fig:interconnection2}, which shows the relative position in the overall network of the five types of games.

The types of games G3 and G4 resulted with longer duration because they consider a deterministic preferential strategy implying displacements to network positions having more neighbors of the same team.
In the specific case of G3, this strategy is adopted only by one of the teams, which ends up implying that the team mostly remaining at the same positions in the overall network, hence the longer duration of the game. The removal of players, leading to the eventual victory of the team following a strategy, takes place mostly as a consequence of the random motion of the team devoid of strategy. Similar game durations have been obtained for G4 in the case of the topologies ER and BA. However, the duration tended to become substantially larger in the case of the GEO type of networks. This result is at consequence of this type of network having a large average topological distance between pairs of nodes, therefore implying in less chance of players of contrary teams meeting each other. Interestingly, the random preferential application of the strategy of G4 adopted in the case of G5 fostered increased mobility of the players, therefore leading to reduced respective duration.

\begin{figure}
  \centering
     \includegraphics[width=1 \textwidth]{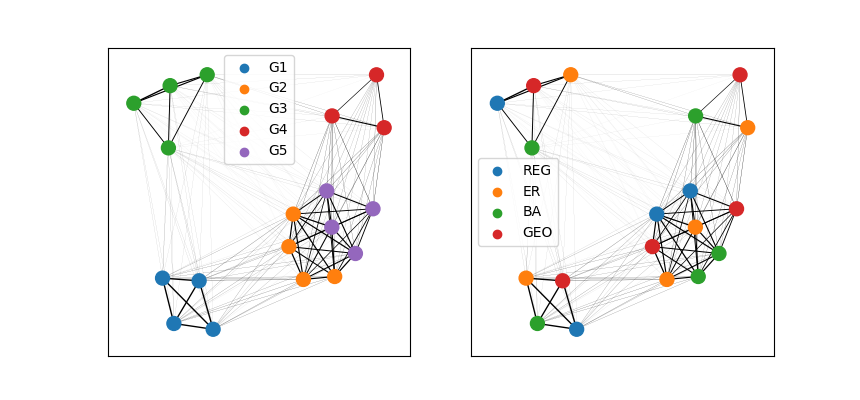} \\
     \hspace{.5cm} (a) \hspace{4.7cm} (b) \\
   \caption{The coincidence similarity network ($D=1$) considering both the percentages of victories/ties and the average and standard deviations of 19 of the considered combinations of games and network topologies. The plots have been identified with labels corresponding to the game types (a) and network topologies (b). Most of the game types resulted well separated one another, except for the G2 and G5 cases.}\label{fig:sim_net4}
\end{figure}

Figure~\ref{fig:sim_net4} depicts the coincidence similarity network obtained by considering both the percentages of victories/ties as well as the average and standard deviation of the duration of 19 combinations of games and topologies (the case G4 in the REG topology has been left out because of its infinite duration).

\begin{figure}
  \centering
     \includegraphics[width=.4 \textwidth]{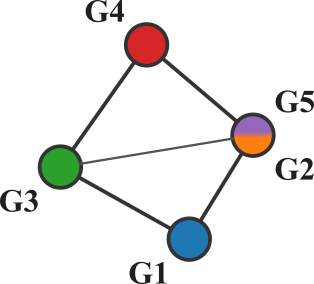} \vspace{0.7cm} \\
   \caption{Summarization of the relative interconnections among the four clusters identified in Fig~\ref{fig:sim_net3}, therefore considering information about \emph{both} the percentages of victories/ties as well as the game durations (averages and standard deviations). Four groups are interrelated in a cyclic manner, as well as the additional relationship between groups G3 and G2/G5.}\label{fig:interconnection3}
\end{figure}

Interestingly, the game types G1, G3, and G4 resulted well separated, indicating their homogeneity when implemented in respective types of complex networks as well as their distinct properties. In addition, the network visualization allowed a comprehensive indication of how the five types of games relate one another. Figure~\ref{fig:interconnection3} summarizes these relationships in terms of the subsumption of the cases respective to each game type.

A \emph{cycle} can be observed involving (in clockwise sequence) the relationships $G1 \leftrightarrow G3 \leftrightarrow G4 \leftrightarrow G5/G2 \leftrightarrow G1$, which incorporates several of the relationships observed considering the percentages of victories/ties and durations separately. A weaker relationship can also be observed between the game types G3 and G2/G5. The effect of the topology on the game types G4 and G5 can be observed from the different widths of the links among the nodes of those respective groups.

The network obtained while taking into account all types of features considered in this work provides an interesting example of situations in which the incorporation of a more complete set of features contributes to a better characterization and grouping of the data elements. However, this is not always the case, since the inclusion of new features can also lead to less definite clustering structures. This can take place in several situations, including when noisy features are incorporated. In addition, the incorporation of several intrinsically related features (redundant one another) can also tend to reduce the clustering in the obtained networks.

\section{Concluding Remarks}

One particularly intriguing aspect underlying complex systems represented by respective networks concerns the possible relationship between diverse types of topologies and dynamics. The present work addressed the subject of relatively simple games taking place on four types of complex networks namely REG, ER, BA, and GEO. In addition to providing the context for exploring the possible relationship between topology and dynamics, the subject of simple games on complex networks is also interesting on itself, especially regarding the identification of simplest possible games that can lead to interesting dynamics obtained by symmetry breaking promoted by specific types of rules governing the respective dynamics. All games in the present work involve parallel, simultaneous movements by both teams.

The first type of game considered here, namely G1, was characterized by involving uniformly random displacements of the players during the games, with pairs of players being removed whenever they meet at a same node. This type of game was characterized by full symmetry of dynamics that implied always a tie between the two teams.

The type of game G2 was derived from G1 by considering the number of neighbors from the same team when two players of distinct teams meet. More specifically, the player having the largest number of neighbors from its own team prevail over the other player, which is therefore removed from the game. This simple rule was found to be enough to break the symmetry underlying game G1, leading to a substantial probability of victory by one of the two teams. Both G1 and G2 have been found to lead to relatively short overall durations. 

The incorporation of strategy was also contemplated in the present work respectively to the types of games G3, G4, and G5. In the case of G3, only one of the teams adopts the strategy of its players deterministically moving to the node with the largest number of team companions, while the other team follows the same rules as in G2. This modification implied in substantially increased probability of the team following the strategy to win the game. At the same time, a substantial increase in the respective game duration has been identified which is a consequence of the players following the strategy to remain close one another at about the same position in the network.

The type of game G4 was then derived from G3 by having both teams to follow in a deterministic manner the above described strategy, which lead both teams having the same probability to win and similar durations in the case of the ER and BA topologies, as well as even longer durations in the case of the GEO networks.
By adopting a randomly preferential application of the strategy instead of the deterministic dynamics adopted in G4, the displacements now allowed increased mobility by both teams, which led to reduced durations (comparable to the cases of G! and G2).

In addition to the specific interesting results summarized above, we also have the verification that the type of network topology tended to have little influence on the respective game dynamics, at least when considered from the perspectives of percentages of victories/ties and game duration, except for the game types G3 and G4 respectively to the BA and GEO topologies. Another interesting results concerns the fact that the probability of victories resulted substantially larger in all cases except G1. Another aspect of special interest regards the fact that the incorporation of deterministic application of the considered strategy led to a substantial increase in the duration of the games, which was promptly reduced with the adoption of preferentially random application of the same strategy. It would be interesting to verify if this is a more general phenomenon or if it is restrict to the considered types of games.

The described approach and results also provided an interesting illustration of how coincidence similarity networks can provide substantial help while understanding and interpreting results, as compared to respective presentations as tables or plots (histograms). In particular, the visualization of the coincidence similarity networks obtained respectively to the probabilities of victory/tie as well as game durations allowed not only the immediate observation of the relationships between each pair of cases but also the identification of possible groups or clusters involving specific types of games or networks. In addition, the coincidence similarity network obtained for the game durations also allowed the identification of the adjacency among the five considered types of games.

The developments and results described in the present work pave the way to several subsequent investigations, including the consideration of other types of networks (e.g.~modular), other types of rules (e.g.~displacements proportional to the degree of the destination node), as well as other additional or alternative strategies (e.g.~considering possible future game configurations).

\section*{Acknowledgments}
Alexandre Benatti thanks MCTI PPI-SOFTEX (TIC 13 DOU 01245.010\\222/2022-44). Luciano da F. Costa thanks CNPq (grant no.~307085/2018-0) and FAPESP (grant 2022/15304-4).

\bibliography{ref}
\bibliographystyle{unsrt}

\end{document}